\documentclass[letter]{aa}

\usepackage{graphicx}
\usepackage{txfonts}
\usepackage{ogonek}
\usepackage{url}

\begin{document} 
 
   \title{Rotation periods of $12\,000$ main-sequence \textit{Kepler} stars: Dependence on stellar spectral type and comparison with $v \sin i$ observations   \thanks{Table 1 are only available in electronic form at the CDS via anonymous ftp to cdsarc.u-strasbg.fr (130.79.128.5) or via http://cdsweb.u-strasbg.fr/cgi-bin/qcat?J/A+A/}}

   \author{M. B. Nielsen
          \inst{1,2}
          \and
	  L. Gizon
          \inst{2,1}
          \and
          H. Schunker
          \inst{2}
          \and
          C. Karoff
          \inst{3}
          }

   \institute{Institut f{\"u}r Astrophysik, Georg-August-Universit{\"a}t G{\"o}ttingen, Friedrich-Hund-Platz 1, 37077 G{\"o}ttingen, Germany\\
             \email{nielsenm@mps.mpg.de}
             \and
	     Max-Planck-Institut f{\"u}r Sonnensystemforschung, Max-Planck-Stra{\ss}e 2, 37191 Katlenburg-Lindau, Germany
	     \and
             Stellar Astrophysics Centre (SAC), Department of Physics and Astronomy, Aarhus University, 
             Ny Munkegade 120, DK-8000 Aarhus C, Denmark
             }

   \date{Received 17 May 2013; accepted 26 July, 2013}

  \abstract
   {}
   {We aim to measure the starspot rotation periods of active stars in the \emph{Kepler} field as a function of spectral type and to extend reliable rotation measurements from F-, G-, and K-type to M-type stars.}
   {Using the Lomb-Scargle periodogram we searched more than 150\,000 stellar light curves for periodic brightness variations. We analyzed periods between 1 and 30 days in eight consecutive \emph{Kepler} quarters, where 30 days is an estimated maximum for the validity of the PDC\_MAP data correction pipeline. We selected stable rotation periods, i.e., periods that do not vary from the median by more than one day in at least six of the eight quarters. We averaged the periods for each stellar spectral class according to B-V color and compared the results to archival $v \sin i$ data, using stellar radii estimates from the \emph{Kepler} Input Catalog.}
   {We report on the stable starspot rotation periods of 12\,151 \emph{Kepler} stars. We find good agreement between starspot velocities and $v \sin i$ data for all F-, G- and early K-type stars. The 795 M-type stars in our sample have a median rotation period of $15.4$\,days. We find an excess of M-type stars with periods less than $7.5$\,days that are potentially fast-rotating and fully convective. Measuring photometric variability in multiple Kepler quarters appears to be a straightforward and reliable way to determine the rotation periods of a large sample of active stars, including late-type stars.} 
   {}

   \titlerunning{Rotation periods of 12\,000 \emph{Kepler} stars}
   \authorrunning{Nielsen et al.}

   \maketitle

\section{Introduction}
Measuring stellar rotation as a function of age and mass is essential to studies of stellar evolution \citep{Maeder2009} and stellar dynamos \citep{BohmVitense2007,Reiners2012b}. 
Rotation can be measured at the surface of individual stars using either spectroscopy \citep[e.g.,][]{Royer2004} or periodic variations in photometric light curves due to the presence of starspots \citep{Mosser2009}. On the Sun, sunspots and plage regions modulate the solar irradiance with periods close to the solar rotation period. Brightness variations are also seen in other stars and are commonly attributed to the presence of magnetic activity in the case of main-sequence cool dwarfs \citep[e.g.][]{Berdyugina2005}. Sunspots and solar active regions have lifetimes of days to weeks (rarely months) \citep{Solanki2003} and are reasonably good tracers of solar surface rotation at low latitudes. Starspots have been observed to persist for even longer periods \citep[e.g.][]{Strassmeier2009}, and they appear at high latitudes as well. 

The \emph{Kepler} mission \citep{Borucki2010} has been monitoring the light emitted by more than $10^5$ stars since its launch in 2009. Treating such an enormous number of stars obviously demands an automated approach. In this paper we present the results of a straightforward automated method for analyzing \emph{Kepler} time series and detecting amplitude modulation due to stellar activity, with the aim of determining stellar rotation periods.
 
\section{Measuring stellar rotation} \label{sec:method}
\subsection{\emph{Kepler} photometry}
We used white light time series with a cadence of $29.42$ minutes from the NASA \emph{Kepler} satellite. The data are released in segments of $\sim90$ days (quarters) through the Mikulski Archive for Space Telescopes\footnote{http://archive.stsci.edu/kepler/} for a total sample of stars currently numbering $\sim 190\,000$. We used quarters 2 through 9 (two years of observations in total). The data was processed for cosmic rays and flat fielding prior to release. We used the version of the data that was corrected by the PDC\_MAP pipeline \citep{Smith2012}. The PDC\_MAP correction attempts to detect and remove systematic trends and instrumental effects, which are common to a large set of adjacent stars on the photometer. In addition, we used the most recent data from the msMAP correction pipeline \citep[where `ms' stands for multi-scale]{Thompson2013} to check for consistency with PDC\_MAP.

From an initial sample of $192\,668$ stars, we discarded targets that are known eclipsing binaries \citep{Matijevic2012}, planet host stars as well as planet candidate host stars, and \emph{Kepler} objects of interest (all lists are available through the MAST portal). We do this to reduce the possibility of false positive detections, since these types of variability may be mistaken for transits of starspots.

\subsection{Detecting rotation periods}
Provided active regions or starspots are present over several rotations of the star, a peak will appear in the periodogram of the time series. Assuming starspots trace surface rotation, this provides a way to measure the rotation rate of the star at the (average) latitude of the starspots. Simulations by \citet{Nielsen2012} show that selecting the peak of maximum power is a suitable method for recovering the stellar rotation period. 

We analyzed the \emph{Kepler} observations as follows:
\begin{enumerate}
\item We compute a Lomb-Scargle (LS) periodogram \citep[see][]{Frandsen1995} for each star in each quarter for periods between $1$ and $100$ days, using PDC\_MAP data.
\item We find the peak of maximum power in this period range and record its period.
\item If the period of the peak falls between $1$ and $30$ days, we consider it due to stellar variability and not instrumental effects.
\item The peak height must be at least four times greater than the white noise estimated from the root mean square (RMS) of the time series \citep{Kjeldsen1995}. 
\end{enumerate}
The lower bound in periods of one day is set to avoid the hot g-mode pulsators with frequencies of a few cycles per day \citep[see][chap. 2]{Aerts2010}. Some contamination from g-mode pulsations is expected for F- or earlier-type stars; however, we have not investigated how to automatically differentiate these pulsations from stellar activity variability. The upper bound in period of $30$ days is the estimated limit for which the PDC\_MAP pipeline does not overcorrect the light curve, to the extent that it completely removes the intrinsic stellar signal \citep{Thompson2013}. We calculated periods up to $100$ days to ensure that any peak found below $30$ days is not a potential side lobe of a dominating long-term trend ($>30$ days). 

\subsection{Selecting stable rotation periods} \label{sec:selection}
Further, we require the that the measured periods are stable over several \emph{Kepler} quarters. Specifically, 
\begin{enumerate}
\setcounter{enumi}{4}
\item we determine the median value of the measured periods over all eight quarters;
\item we select stars for which the median absolute deviation (MAD) of the measured periods is less than one day, i.e., $\mathrm{MAD} < 1$\,day \label{pt:MADlim}; The MAD is defined as $\mathrm{MAD} = \left\langle {\left| {P_i - \left\langle {P_i } \right\rangle } \right|} \right\rangle $ (where $\left\langle {} \right\rangle$ is the median).
\item From these, we select stars with six or more (out of eight) measured periods within $2$ MAD of the median period \label{pt:MADrange};
\item we repeat this method using the msMAP data and flag stars that do not satisfy the above criteria.
\end{enumerate}

\begin{figure}
	\centering
	\includegraphics[width=1\columnwidth,keepaspectratio,trim = 0 0 0 20, clip = True]{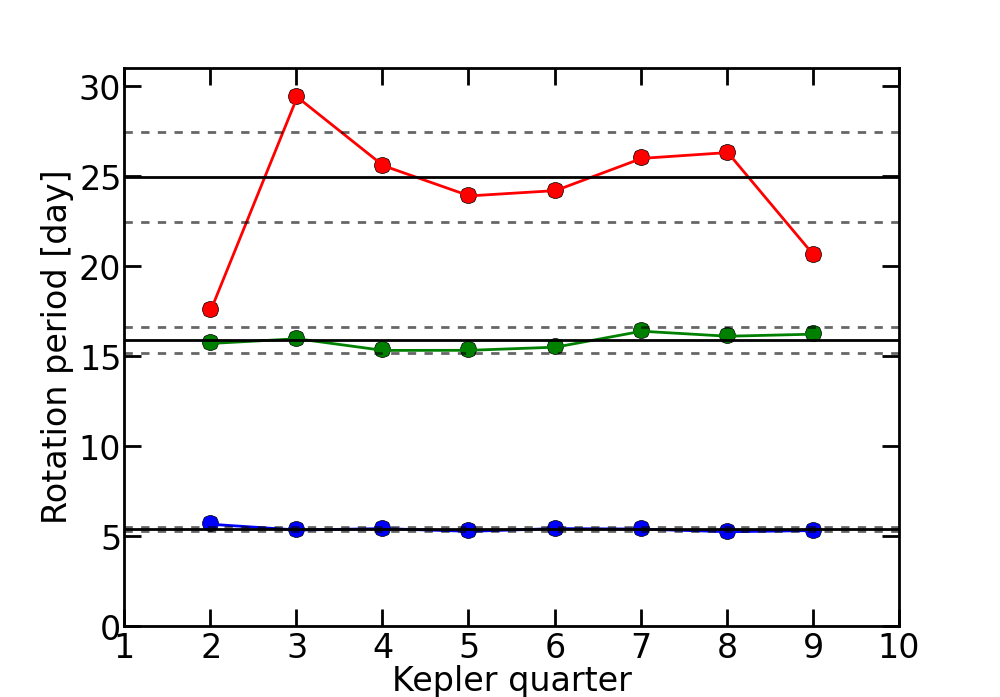}	
\caption{Measured periods in each quarter of observation for three \emph{Kepler} stars (red, green and blue). The solid line is the median period over all quarters for each star. The dashed horizontal lines indicate two median absolute deviations (MAD) from the median period. The long period target (red) is discarded by the algorithm due to high scatter in the period measurements ($\mathrm{MAD} = 1.26$ days). The green and blue target are examples of stars that meet the selection criteria.}
\label{fig:P_v_Q}
\end{figure}

We use the MAD since it is less sensitive to outliers than the standard deviation \citep{Hoaglin2000}. Requiring that a particular variation for a star is visible in multiple quarters reduces the risk of the detection coming from low-frequency noise from, say, instrumental effects. The LS periodogram is calculated at $\sim1300$ linearly spaced frequencies between $1.2 \times 10^{-2}$~mHz and $3.9 \times 10^{-4}$~mHz ($0.03 \, \mathrm{d}^{-1}$ and $1 \, \mathrm{d}^{-1}$). The MAD limit of one day (point \ref{pt:MADlim}), along with the signal attenuation introduced by the PDC\_MAP correction, leads to a selection bias towards stars with shorter rotation periods.  Examples of periods detected for three stars are shown in Fig.~\ref{fig:P_v_Q}. 

We applied the scaling relation by \citet{Kjeldsen2011} to find timescales for p-mode pulsations in cool main sequence stars and red giants. We found that stars with $\log{g} \lesssim 2$ have pulsation periods that can potentially overlap with the range investigated in this work. We opted for a conservative approach and discarded stars with $\log{g} < 3.4$ to remove red giants from the sample. Following the scaling relation, the main sequence stars were found to have pulsation timescales from minutes to hours, far below our lower period limit. Once all the above criteria are met, the rotation period, $P_\mathrm{rot}$, is defined as the median of the valid periods. This selection process leaves us with 12\,151 stars out of the original sample of 192\,668. When using the msMAP data, we find that $\sim80\%$ of these stars satisfy the above criteria as well. Of these, $0.9\%$ differ from the PDC\_MAP results by more than one resolution element, predominantly because the msMAP data shows the first harmonic instead of the fundamental period of the variability. The msMAP pipeline treats the long periods ($P_{\mathrm{rot}} \lesssim{15}$ days) differently than the PDC\_MAP \citep[see][]{Thompson2013}. %These stars are not removed but simply 

The results for all 12\,151 stars are shown in Table 1, which is provided as online material through the CDS. Column 1 gives the \emph{Kepler} Input Catalog (KIC) name of the star, cols. 2 and 3 are the rotation period and scatter (MAD), and cols. 4 to 8 give the $g-r$ color, $E(B-V)$, radius, $\log{g}$, and $T_{\mathrm{eff}}$, respectively, all of which are KIC values. Column 9 is a flag indicating whether each msMAP-corrected data set satisfies the criteria of section \ref{sec:method}. Column 10 gives the msMAP period of the stars where we find rotation rates from the two data sets that differ by more than one resolution element. 

\section{Consistency with $v \sin i$ measurements} \label{sec:results}
We performed a rudimentary spectral classification of the stars based on their $B-V$ color indices. The KIC provides $g-r$ values that we converted to $B-V$ using the relation $B - V = 0.98 \left( {g - r} \right) + 0.22$ given by \citet{Jester2005} for stars with R-I color indices $< 1.15$. The $B-V$ values are dereddened using the $E\left( {B - V} \right)$ values from the KIC \citep[see][for details on their derivation]{Brown2011}. The calculated $B-V$ colors are mapped to a spectral type as per \citet[][Appendix B]{Gray2005}. 

We compared the periods found in this work with $v \sin i$ values compiled by \citet{Glebocki2005}. This list contains $v \sin i$ values and spectral types for $\sim 30\,000$ cluster stars almost isotropically distributed in galactic coordinates. Using our approximate spectral classification we compared the median equatorial velocities of this sample with our results from the \emph{Kepler} targets. From the $v \sin i$ sample we select stars with apparent V magnitude from $6$ to $15$, roughly equivalent to the magnitude range of the \emph{Kepler} targets. We discard any stars from the $v \sin i$ sample that have been ambiguously labeled as `uncertain'. Lastly, we select only dwarf stars since this is the main constituent of the stars selected by our method, reducing the $v \sin i$ sample to be comparable in number to our \emph{Kepler} target list.

For each spectral type (s.t.) in our sample of \emph{Kepler} targets we calculate the median equatorial rotational velocity by
$ \bar{v}({\rm s.t.}) = 2\pi \left\langle {R_{\mathrm{KIC}} / P_{\mathrm{rot}}} \right\rangle $ where $\left\langle {} \right\rangle $ denotes the median over stars with spectral type s.t., $R_\mathrm{KIC}$ is the KIC stellar radius, and $P_\mathrm{rot}$ is the rotation period determined by our algorithm. The KIC radii are notoriously bad estimates in some cases; however, by comparing these with characteristic radii given in \citet[][Appendix B]{Gray2005} we find that the median values in the region of F0 to K0 agree within $\sim15\%$. Outside this range of spectral types the radii are initially overestimated, but become strongly underestimated for A-type stars and earlier.

Since we have such a large statistical sample of $v \sin i$ measurements within each spectral type, a random distribution of inclinations of rotation axes gives $\left\langle \sin i \right\rangle = \pi /4$ \citep[e.g.,][]{Gray2005}. Therefore the median equatorial velocity for a given spectral type can be approximated by $(4/\pi) \left\langle v \sin i \right\rangle $, which is directly comparable to the $\bar{v}(s.t.)$ computed for the \emph{Kepler} targets.

\begin{figure}
	\centering
	\includegraphics[width=1\columnwidth,keepaspectratio,trim = 0 0 0 90, clip = True]{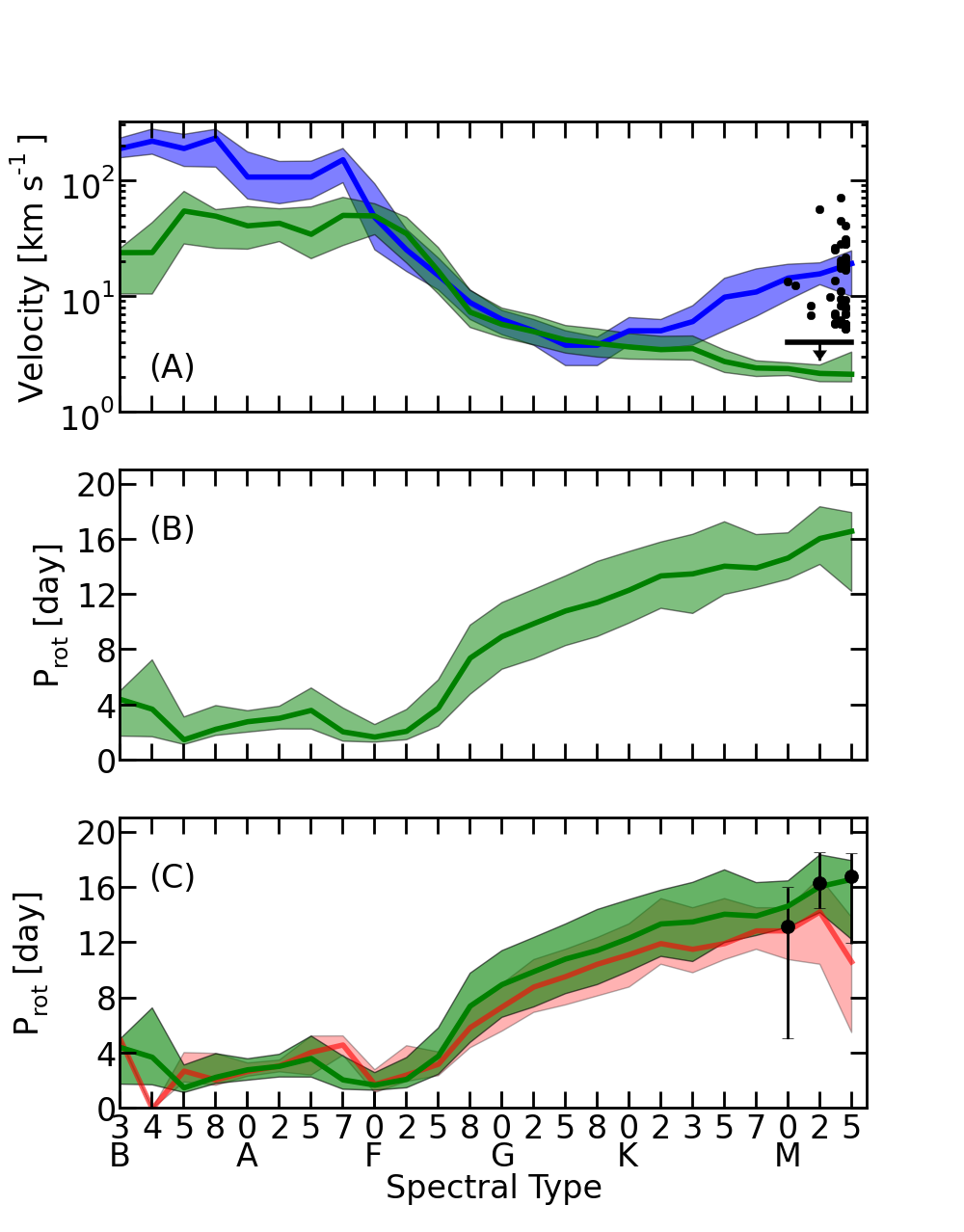}
\caption{Panel A: The blue curve is the median equatorial velocity $(4/\pi) \left\langle v \sin i \right\rangle$ for each spectral type from \citet{Glebocki2005}. The green curve shows the equatorial velocity of the \emph{Kepler} targets, $\bar{v}({\rm s.t.})$, derived from the measured rotation periods and the KIC radii. The black points show measurements by \citet{Reiners2012c}. In this sample 201 stars have an upper $v \sin i$ limit of $4\,\mathrm{km/s}$ (due to instrumental limitations), these stars are represented by the solid bar. Panel B: The rotation periods $P_{\rm rot}$ of the stars in our sample, averaged within each spectral type. Panel C: The same as panel B, but for comparison we show the median of the rotation periods measured by \citet{McQuillan2013} (black points with errorbars), for the stars overlapping with our sample. Similarly, the red curve shows the median of the rotation periods found by \citet{Debosscher2011}. Shaded areas and error bars span the upper and lower 34$^\mathrm{th}$ percentile values from the median.}
	\label{fig:veq_v_spectral_type}
\end{figure}

Panel A in Fig.~\ref{fig:veq_v_spectral_type} shows clear agreement, from late A-type to early G-type stars, between the median equatorial velocities of the $v \sin i$ sample and those we derived from our measured \emph{Kepler} rotation periods. For additional comparison we included the $v \sin i$ measurements from \citet{Reiners2012c}, where the horizontal bar represents 201 stars with velocities below $4\,\mathrm{km/s}$. Panel B shows our measured rotation periods $P_{\mathrm{rot}}$, as a function of spectral type, which we used to calculate the equatorial velocities. We also calculated the median periods of the \emph{Kepler} stars found in \citet{Debosscher2011} and \citet{McQuillan2013} that are present in our sample. These are shown in comparison to our results in panel C. We find that $\sim96\%$ and $\sim97\%$ of their measured rotation rates fall within one frequency resolution element ($1/90 \, \mathrm{d}^{-1}$) of our values for the corresponding stars. 

\section{Rotation of late type stars}
Owing to their small radius and long rotation period, the $v \sin i$ measurements of late type stars are limited by spectral resolution. The $v \sin i$ measurements of \citet{Reiners2012c} have a lower limit at $\sim4\, \mathrm{km/s}$. The KIC radii are overestimated for stars later than K0 so our velocities are therefore upper limits, but we still systematically find rotation slower than $\sim4\, \mathrm{km/s}$. This places our measurements at or below the lower limit for spectroscopically determining rotation for these types of stars. 

\begin{figure}
	\centering
	\includegraphics[width=1\columnwidth,keepaspectratio,trim = 0 0 0 20, clip = True]{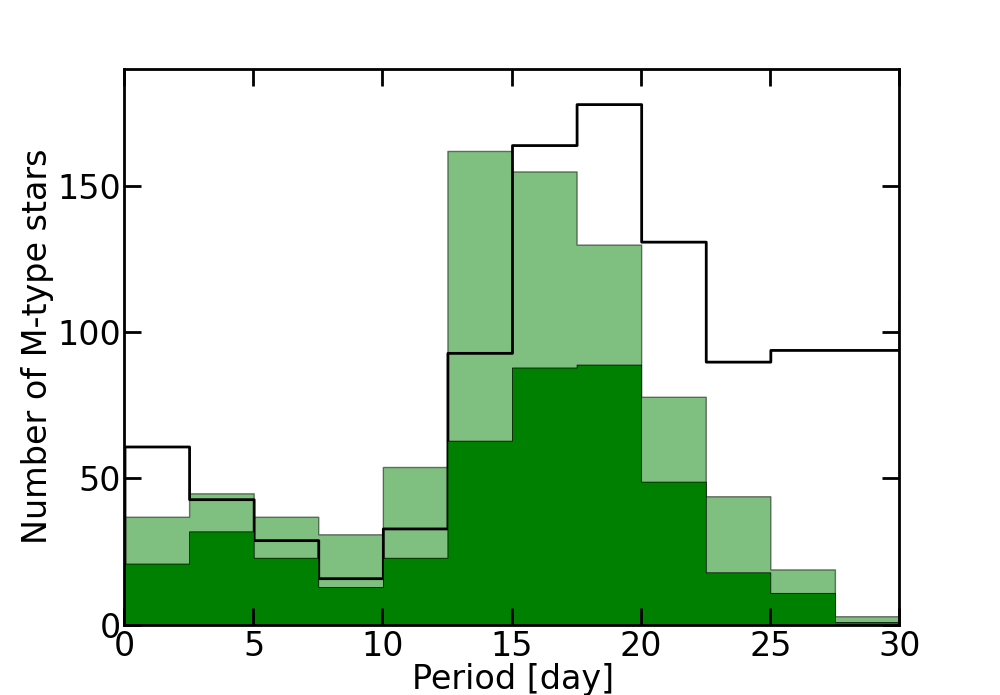}
\caption{The light green histogram shows the distribution of rotation periods for the 795 M-type stars in our sample (median, 15.4 days). For comparison the black line shows the results by \citet{McQuillan2013} for periods less than 30 days. The dark-green histogram shows the distribution of rotation periods measured by \citet{McQuillan2013} for the stars in common with our sample.}
	\label{fig:coolstars}
\end{figure}

The late M-type stars are of particular interest since they represent the transition from solar-like convective envelopes to fully convective interiors. Figure~\ref{fig:coolstars} shows the distributions of stars that have been classified as M-type, found in this work and those by \citet{McQuillan2013}. The median of the distribution in our sample is $15.4$ days. There appears to be good agreement with the result of \citet{McQuillan2013}. We note an excess of fast rotators with $P < 7.5$\,days. The study by \citet{Reiners2012c} shows that the number of magnetically active M-type dwarfs increases approximately after spectral subtype M3, which appears correlated with an increase in the number of fast rotators (see top panel of Fig.~\ref{fig:veq_v_spectral_type}), likely marking the transition to full convection. Our sample contains 795 M-type stars, where $\sim15 \%$ have periods less than 7.5 days; i.e., these are potentially fast-rotating, fully convective M-type dwarfs, such as those found by \citet{Reiners2012c}. As a result of the KIC $\log{g}$ having errors of approximately $\pm0.4$\,dex \citep{Brown2011}, the sample probably still contains some cool giants. A visual inspection of the power spectra of the M-type stars, with periods $P < 7.5$\,days, indicated that 3 out of 119 stars had p-mode pulsations that could potentially be misidentified as periodic activity. Thus the contamination by p-mode pulsations appears to be negligible. A more rigorous determination of $\log{g}$ and spectral type for these stars is required to distinguish stellar evolutionary stages. However, even a fraction of our sample of M-type dwarfs still significantly complements the existing literature on this critical region for understanding differences in stellar dynamos.
 
\section{Conclusions} \label{sec:conclusion}
We developed a straightforward, automated method for detecting stable rotation periods for a large sample of \emph{Kepler} targets. A total of 12\,151 stars ranging from spectral type B3 to M5 all show recurring periods in at least six of the eight quarters of analyzed \emph{Kepler} data. This, along with the requirement that the stars in question have a maximum MAD of one day, are empirically determined criteria. Using the KIC stellar radii, we found very good agreement between the equatorial rotational velocities derived from the \emph{Kepler} rotation periods and independent $v \sin i$ measurements for F0 to K0 type stars. For later type stars we find an inconsistency with the \citet{Glebocki2005} catalog. This is due to an age difference between the two samples, since the \citet{Glebocki2005} catalog mainly consists of young open cluster stars. However, our results for M-type stars agree well with those reported by \citet{Reiners2012c} and \citet{McQuillan2013}. The study by \citet{Debosscher2011} analyzed a different set of \emph{Kepler} data, but we still find good agreement with the stars overlapping with our sample.

We have primarily studied stars that have long-lived stellar activity signatures. We tested our method on solar-disk integrated light at solar maximum (January 2001 to March 2003), observed by the VIRGO instrument aboard SOHO \citep{Froehlich1997}. The VIRGO green channel data were divided into 90-day segments, and we applied the method described in Sect.~\ref{sec:method}. The stability criterion was not met due to a period scatter of $\mathrm{MAD} = 3.7$\,days. Analysis of other segments of VIRGO data from 1995 to 2013 also resulted in rejection. This shows that stars with solar-like (low) activity are rejected from our sample; for such stars, rotation period measurements are too noisy and not stable enough over time. A further bias is the upper limit of $P\leq30$\,days chosen because the PDC\_MAP does not yield reliable corrections for longer periods. An obvious improvement to our analysis would be to investigate ways of increasing this upper limit, for a complete view of slow rotators.

As expected, we found that hot stars rotate faster than their cooler counterparts \citep{Barnes2003,Kraft1970}. The rotation periods of hot stars should be treated with some caution since they may be false positives from g-mode pulsations. Nevertheless, our analysis method detects periodic variations in brightness that satisfy the selection criteria. The measured variability is not necessarily of magnetic origin, but could arise from, for example, chemical surface inhomogeneities in chemically peculiar hot stars. Chemical spots can produce photometric variability that traces the stellar rotation \citep{Wraight2012,Paunzen2013}. The short periods measured for the early spectral types (see Fig. \ref{fig:veq_v_spectral_type}) in our sample are therefore not contradictory to the expected fast rotation of hot stars \citep{Royer2004}. 
 
This work is only one step toward characterizing of the rotation of stars in the \emph{Kepler} field. \emph{Kepler} photometry is proving very useful in adequately sampling the slow rotation rates of cool, faint stars. Future work will include detailed starspot modeling in order to measure latitudinal differential rotation \citep{Reinhold2013} and asteroseismology \citep[e.g.,][]{Deheuvels2012} to infer internal differential rotation. The \emph{Kepler} observations offer unique possibilities for calibrating the mass-age-color relations in gyrochronology \citep{Skumanich1972, Barnes2007} and exploring the close relation between stellar rotation and activity cycles.

\begin{acknowledgements}
M.N., L.G., and H.S. acknowledge research funding by the Deutsche Forschungsgemeinschaft (DFG) under grant SFB 963/1 ``Astrophysical flow instabilities and turbulence'' (Project A18, WP ``Seismology of magnetic activity''). Funding for the Stellar Astrophysics Centre is provided by The Danish National Research Foundation. The research is supported by the ASTERISK project (ASTERoseismic Investigations with SONG and Kepler) funded by the European Research Council (Grant agreement no.: 267864). This paper includes data collected by the Kepler mission. Funding for the Kepler mission is provided by the NASA Science Mission directorate. Some/all of the data presented in this paper were obtained from the Mikulski Archive for Space Telescopes (MAST). STScI is operated by the Association of Universities for Research in Astronomy, Inc., under NASA contract NAS5-26555. Support for MAST for non-HST data is provided by the NASA Office of Space Science via grant NNX09AF08G and by other grants and contracts. The authors thank Ansgar Reiners for helpful discussions.

\end{acknowledgements}
\bibliographystyle{aa.bst}
\bibliography{rotation_paper.bib}
\end{document}